\documentclass[%
 reprint,
superscriptaddress,
showpacs,preprintnumbers,
 amsmath,amssymb,
 aps,
]{revtex4-1}

\usepackage[utf8]{inputenc}
\usepackage[english]{babel}
\usepackage{amssymb}
\usepackage{bm}
\usepackage[dvipsnames]{xcolor}
\usepackage{bbm}
\usepackage{bbold}

\usepackage{latexsym}
\usepackage{amsmath}
\usepackage{float}

\usepackage{soul}
\usepackage{ulem}
\usepackage{graphicx}
\usepackage{dcolumn}
\usepackage{bm}
\usepackage{color}
\usepackage{natbib}
\usepackage[colorlinks=true,citecolor=blue]{hyperref}
\hypersetup{colorlinks=true,citecolor=blue,linkcolor=red
,urlcolor=blue}

\begin{document}
\date{\today}

\title{RKKY plateau in zero- and one-dimensional triangular Kagome lattice models  }
\author{Moslem Zare }
\email{mzare@yu.ac.ir}
\affiliation{Department of Physics, Yasouj University, Yasouj,75914-353, Iran}

\date{\today}

\begin{abstract}
Motivated by the research interests on realizing flat bands and magnetization plateaus in kagome lattices, we study the electronic properties and magnetic interactions in both zero- and one-dimensional triangular Kagome lattice (1D-TKL) models, by using the real-space Green’s function approach in tight-binding model. We firstly study the electronic properties of both 0D and 1D-TKL in the presence of staggered sublattice potential, then, by analyzing the  Ruderman-Kittel-Kasuya-Yoshida (RKKY) interaction in these lattice structures, the magnetic ground states of both 0D and 1D-TKL in the presence of two magnetic adatoms are evaluated.
It is found that the 1D channels of TKL show different electronic and magnetic behaviors due to different values of the hopping integrals and spin-orbit couplings. Two important salient features of these TKLs are the presence of flat bands in their band structure as well as the emergence of the RKKY plateaus versus the Fermi energy.
These RKKY plateaus have not been reported before, to the best of our knowledge.
The most remarkable observation is the potential and Fermi energy variation of the width and location of the RKKY plateaus in both 0D and 1D-TKLs.
The spatial configurations of the magnetic impurities can also dramatically change the quality and quantity of the RKKY plateaus.
We believe that our results provide significant insights towards designing further experiments to search for the realizing of the flat bands and magnetization plateau phases in spintronics and pseudospin electronics devices based on TKLs.

\end{abstract}

\maketitle

\section{Introduction}\label{sec:intro}
In the past few years, the research on two dimensional (2D) graphene-like materials has extended to other structures with honeycomb arrangements, largely fueled by the search for exotic states of matter.

Recently, a class of layered materials called as “kagome lattices”, have been the subject of numerous studies in condensed matter physics due to their potential remarkable technological applications.
A kagome lattice, as a classical spin liquids~\cite{Garanin99}, consists of triangular and hexagonal motifs in a network of corner-sharing triangles. It can be realized by cold atoms in the optical lattices~\cite{Duanprl03,Panahi2011,Gemelke2009,Y.H.Chen10,Jo2012}, the geometrically frustrated lattice of the Cu spins in triangular kagome lattice (TKL) compound $Cu9_Cl_2(cpa)_6. nH_2O $~\cite{Maruti94,Mekata98}, a bilayer form of MgB$_6$ sandwiching the Mg layer between two B$_{3}$ kagome layers~\cite{Xie2015}, the self-assembled metal-organic molecules~\cite{Mao2009}, and the kagome layers of pyrochlore oxides~\cite{Gardner2010}.

The triangular kagome lattice provide a promising avenue for realizing exotic quantum phenomena such as frustrated quantum antiferromagnet~\cite{Waldtmann98}, frustrated magnetic ordering\cite{Balents2010,Nisoli2013}, quantum spin liquid~\cite{Norman1990}, ferromagnetism~\cite{Mielke1992,Tanaka2003}, and topologically non-trivial states\cite{Ohgushi2000,Guo2009,Tang2011,Guterding2016}.

It has long been known that, antiferromagnetic Ising model on a kagome lattice does not order~\cite{Syozi1951}.
Due to the high degeneracy of their ground state and ensuing large fluctuations, classical antiferromagnets on kagome and pyrochlore lattices are frustrated systems~\cite{Reimers91}.

Magnetization plateaus, a family of known non-trivial quantum effects in the condensed matter physics, have been observed experimentally in the kagome lattices~\cite{DMRG2,KN53,Hida2001,Schulenburg2002,Richter2004,Honecker2004,Damle2006}.
Numerous theoretical studies have been performed to explain the magnetization plateaus in various magnetic materials, including: non-homogeneous Heisenberg chains~\cite{Hida,pl4}, spin-ladders \cite{ladder}, spin-1/2 onedimensional (1D) Heisenberg model with a kagome strip~\cite{Morita2018}, crystallization of the magnetic particles~\cite{pl1,KM} and some two dimensional models \cite{2d,Hon}.
It has long been known that the purely quantum phenomena are responsible for the magnetization plateaus mechanisms in such structures.
In spite of many experimental and theoretical studies, unfortunately, the magnetic exchange coupling, the main focus of this article, is still not understood.

On the other hand, however, the flat bands corresponds to highly
degenerate ground-state manifolds and strictly dispersionless in the whole Brillouin zone in the kagome networks ~\cite{Sutherland1986,Aoki1993,Matsumura1996,Tasaki1998,Yoshino2004,Balents2008,Vanmaekelbergh2014,Desyatnikov2014,Schmidt2015,Lai2015,Flach2016,Vicencio2016,Flach2017o,Flach2017c,Aoki2017,Flach2018r,Schmelcher2018}, have become a subject undergoing intense study in the field of strong correlation physics due to the complete quenching of the kinetic energy~\cite{Lieb1989,Sarma2007,Cooper2012,Zhang2013,Hausler2015,Derzhko2015,Volovik2016,Torma2016,Nguyen2018,Li2018,Wen2011,Sarma2011,Mudry2011,Sheng2011,Bernevig2011,Ran2011,Franz2012,Bergholtz2012,Sarma2012,Lauchli2012,Liu2013a,Liu2013b,Udagawa2017,TaoLi2018,Tovmasyan2016}.
and provide a platform for pseudospin electronics~\cite{Zhang2016a,A.CarvalhoEPL,H.Guo14}.

Several meaningful predictions such as the Wigner crystallization in the honeycomb lattice~\cite{Sarma2007}, the nontrivial conductivity behavior in the presence of long-range Coulomb interactions~\cite{Hausler2015}, and the huge critical temperature for the superconductivity~\cite{Volovik2016} have been proposed in the flat band systems.
As a promising venue to realize fractional Chern insulators, a nearly flat band with a finite Chern number, was recently proposed partially analogous to the case of the flat Landau level~\cite{Wen2011,Sarma2011,Mudry2011,Sheng2011,Bernevig2011,Ran2011,Franz2012,Bergholtz2012,Sarma2012,Lauchli2012,Liu2013a,Liu2013b}.
Interestingly, the experimental realization of nearly flat bands in conventional solid state systems e.g., the photonic crystals~\cite{Thomson2015,Amo2018, Chen2016a, Chen2016b}, optical lattices~\cite{Manninen2013,Takahashi2015,Bloch2016,Takahashi2017}, manipulated atomic lattices~\cite{Liljeroth2017,Swart2017}, and various metamaterials~\cite{Kitano2012,Kitano2016} were also reported.
It was pointed out that the presence of almost flat bands, in the twisted bilayer graphene at magic angle, must be the principal origin of the Mott insulating phases and the associated superconductivity~\cite{Herrero2018a,Herrero2018b}.
More recently a nearly dispersionless band is detected in the layered Fe$_3$Sn$_2$, by ARPES measurements~\cite{Zhang2018}.

Designing the lattice structures which produce the flat band at Fermi energy has attracted much attention recently because of its potential
applications in nanoelectronics, and magnetoectronics. The presence of flat bands at Fermi energy gives rise to the large density of states and is responsible for the flat band ferromagnetism~\cite{K.Nakada,E.Lieb,K.Kusakabe,M.Fujita96}. There are primarily ways toward creating flat bands in nanoribbons~\cite{K.Nakada,E.Lieb,K.Kusakabe,M.Fujita96}. The modification of zigzag edge by attaching Klein’s bonds gives rises to the partial flat band in graphene nanoribbons.

The nonequality between the sublattices in bipartite lattices is one of simple ways to obtain N-degenerated flat bands at the Fermi energy with $N=|N_A-N_B|$, where $N_A$ and $N_B$ are the number of A and B-sublattices, respectively~\cite{E.Lieb,J.Fernandez,EzawaPhysicaE,H.Tamura02}.
As is well known, the flat bands have a key role in the ferromagnetism of the triangular kagome lattice, that induces the quantum anomalous Hall effect in the presence of spin-orbit coupling~\cite{S.Kim17scirept}.
Up to now, a kagome network remains a classic problem in the field of highly frustrated magnetism and despite a number of numerical~\cite{Lauchli2011,Gotze2011} and variational investigations~\cite{Hermele2008,Iqbal2011}, the spin-1/2 Heisenberg model on the kagome lattice in zero field is an open problem that is still not fully understood.

Motivated by recent theoretical and experimental studies of realizing flat bands and magnetization plateaus in kagome lattices, in this work we have addressed the problem of indirect exchange coupling in zero and one-dimensional boron triangular kagome lattices, in the presence of staggered potential.
The electronic properties of the BTKLs are also reported that resulted in establishing a unified picture of the nature of indirect exchange interaction, known as Ruderman-Kittel-Kasuya-Yosida (RKKY) interaction~\cite{Ruderman, Kasuya, Yosida} mediated by a background of conduction electrons of the host material, BTKL.

We have recently addressed the problem of isolated magnetic adatoms placed on silicene~\cite{moslem-si2} and phosphorene sheet~\cite{MoslemBP1} as well as on zigzag silicene nanoribbons~\cite{moslem-si1}, bilayer phosphorene nanoribbons~\cite{MoslemBP2} and both zigzag and armchair B$_2$S nanoribbons~\cite{MoslemBS1,MoslemBS2}.

Motivated by the future potential of the magnetic kagome lattices, in this work we have addressed the problem of indirect exchange coupling in zero and one-dimensional boron triangular kagome lattices, named as B$_{9}$-KL, in the presence of staggered sublattice potential.
Within the tight-binding (TB) model we exploit the Green's function formalism (GF) to reveal the RKKY interaction between two magnetic impurities placed on a BTKL.
Using the newly developed tight-binding model for BTKL we show the existence of various RKKY plateaus for certain range of Fermi energies in the Heisenberg-like RKKY model. To the best of our knowledge, it is the first time that the RKKY plateau is systematically reported and as we will see this idea has been advocated forcefully.

The remainder of the manuscript is organized as follows: In Sec.~\ref{sec:intro}, we describe the systems under consideration, i.e., zero and one-dimensional boron triangular kagome lattices. In Sec.~\ref{sec:theory}, we discuss our numerical results of the exchange plateaus and electronic properties. To do so, a tight-binding model Hamiltonian for 2D boron triangular kagome lattice is presented and then the band spectrum of it is investigated then the theoretical framework which will be used in calculating the RKKY interaction is introduced, from the real space Green’s function. Finally, a summary is given in Sec.~\ref{sec:summary}.

\section{Theory and model}\label{sec:theory}
In this section, we outline our formalism to obtain the magnetic ground state of both 0D and 1D BTKL obtained from the real space Green’s function approach and clarify their electronic properties.
To study indirect magnetic coupling between two localized magnetic moments embedded in BTKL, we consider the indirect exchange coupling between magnetic impurities to be of the RKKY form.

Using a second-order perturbation~\cite{Ruderman,Kasuya,Yosida}, the effective magnetic interaction between magnetic moments ${\bf S}_1$ and ${\bf S}_2$ at positions ${\bf r}$ and ${\bf r'}$, induced by the quantum effects from the free carrier spin polarization, reads as

\begin{equation}
E({\bf r},{\bf r'}) = J ({\bf r},{\bf r'})  {\bf S}_1 \cdot {\bf S}_2,
\label{RKKYE}
\end{equation}

The RKKY interaction $J ({\bf r},{\bf r'}) $ is explained using the static carrier susceptibility as follows
\begin{equation}
J ({\bf r},{\bf r'}) = \frac{\lambda^2 \hbar^2 }{4} \chi ({\bf r},{\bf r'}),
\label{RKKYJ}
\end{equation}
where $\lambda$ is the contact potential between the impurity spins and the itinerant carriers and the static spin susceptibility $ \chi ({\bf r},{\bf r'})$ can be written in terms of the integral over the unperturbed Green's function $G^0 $ as
\begin{equation}
\chi ({\bf r},{\bf r'}) =
- \frac{2}{\pi} \int^{\varepsilon_F}_{-\infty} d\varepsilon \
{\rm Im} [G^0 ({\bf r}, {\bf r'}, \varepsilon) G^0 ({\bf r'},{\bf r}, \varepsilon)],
\label{chiGG}
\end{equation}
where $\varepsilon_F$ is the Fermi energy. The expression for the susceptibility may be obtained by using the spectral representation of the Green's function
\begin{equation}
G^0 ({\bf r},{\bf r'},\varepsilon)= \sum_{n,s} \frac{\psi_{n,s}({\bf r})\psi^{*}_{n,s}({\bf r'})}{\varepsilon+i\eta - \varepsilon_{n,s}},
\label{GFspct}
\end{equation}
where $\psi_{n,s}$ is the sublattice component of the unperturbed eigenfunction with the corresponding energy $\varepsilon_{n,s}$. For a crystalline structure, ${n,s}$ denotes the band index and spin, respectively.
Substituting Eq. (\ref{GFspct}) into Eq. (\ref{chiGG}), finally, the effective strength for the RKKY interaction is obtained from the energy integration of the two-particle Green’s function.
The analytical background of this approach has been already presented in details~\cite{moslem-si1,MoslemBP2} which is not discussed here. From those analytical reports, the RKKY interaction can be expressed in the following desired result

\begin{eqnarray}
\chi({\bf r},{\bf r'}) &&=2 \sum_{\substack{n,,s \\ n',s'}}[ \frac{f(\varepsilon_{n,s})-f(\varepsilon_{n',s'})}{\varepsilon_{n,s}-\varepsilon_{n',s'}}\nonumber\\
&&\times \psi_{n,s}({\bf r})\psi^{* }_{n,s}({\bf r'})\psi_{n',s'}({\bf r'})\psi^{*}_{{ n'}s'}({\bf r})].
\label{chiE}
\end{eqnarray}
where, $f(\varepsilon)$, is the Fermi function.
This is a well-known formula in the linear response theory that is the main equation in this work.

\subsection{1D boron kagome lattice}\label{subsec.TBnlayer}
In this section, firstly we restrict ourselves to general geometrical description of both 0D and 1D BTKL and then, since the indirect exchange interaction between magnetic moments is significantly affected by the electronic structure of the host material, here BTKL, the electronic structures of the 0D-BTKL, known as BTKL quantum dot, and 1D nanoribbons of the BTKL are studied using tight-binding model.

A comparatively recent addition to the 2D kagome family, named as boron triangular kagome lattice, is one of the many 2D polymorphs of monolayer boron.
Boron, the left neighbor of carbon in the periodic table, on the one hand has only three valence electrons (it has one electron less than carbon), which would favour metallicity, but on the other hand its valence electrons are sufficiently localized that insulating states emerge.
Because of this frustration (situated between metals and insulators in the periodic table) boron exhibits a rich variety of structural complexities such as zero-dimensional all-boron fullerene-like cage cluster $B_{40}$~\cite{mannix2015synthesis,Gonzalez2007prl,Gonzalez2007nrl}, 1D boron nanowires and nanotubes~\cite{Otten2002,F.Liu-nanowire,Ciuparu2004,F.Liu2010}, double-ring tubular structures~\cite{Kiran2005,Oger2007,An2006}, graphene-like 2D structure, known as borophene~\cite{H.Liu13,X.Wu12,Y.Liu13,Z.Zhang15} and 3D superhard boron phases~\cite{Eremets2001}.
However, this subtle balance between metallic and insulating states is simply changed by temperature, pressure and impurities~\cite{Oganov2009}.
In regard to the 2D metal-boron lattices such as MoB$_{4}$~\cite{Xie2014}, FeB$_{2}$~\cite{Zhang2016a}, TiB$_{2}$~\cite{Zhang2014}, FeB$_{6}$~\cite{Zhang2016} and MnB$_{6}$~\cite{Li2016}.

Despite a number of theoretical attempts to predict 2D boron allotropes, only a few 2D boron sheets have been synthesized on metal substrates, such as a quasi-2D layer of $\gamma$-B$_{28}$~\cite{Tai2015}, a 2D triangular sheet (generally referred to as borophene)~\cite{Mannix2015}, and borophene with stripe-patterned vacancies~\cite{Feng2016}.
The excessive electrons occupying the antibonding levels results in a fully planar-unbuckled stable state for triangular boron lattice~\cite{Tang2007}.

Very recently in a comprehensive calculations based on the first-principles evolutionary materials design~\cite{S.Kim17scirept}, Kim et al. investigated possibilities for 2D boron phases with the Mg atoms as guest atoms on a silver substrate.
In this work, a stable form of the boron triangular kagome lattice, named as B$_{9}$-KL composed of triangles in triangles on a 2D sheet, has been predicted and some exotic electronic properties, such as topologically non-trivial flat band near the Fermi energy, half-metallic ferromagnetism, and quantum anomalous Hall effect in the presence of spin-orbit coupling reported for this form of the BTKL.
The high temperature dynamical stability of B$_{9}$-KL, verified by calculating the full phonon spectra and performing first-principles calculations, shows that 2D B$_{9}$-KL should be sufficiently stable at room temperature and higher temperatures.

However, B$_{3}$ and B$_{9}$-$t$KL (called a twisted kagome lattice) are another allotropes of 2D boron kagome sheets that are energetically less stable than the B$_{9}$-KL and are known to be dynamically unstable due to the lack of electrons~\cite{Xie2015}.

\begin{figure*}
\includegraphics[width=15cm]{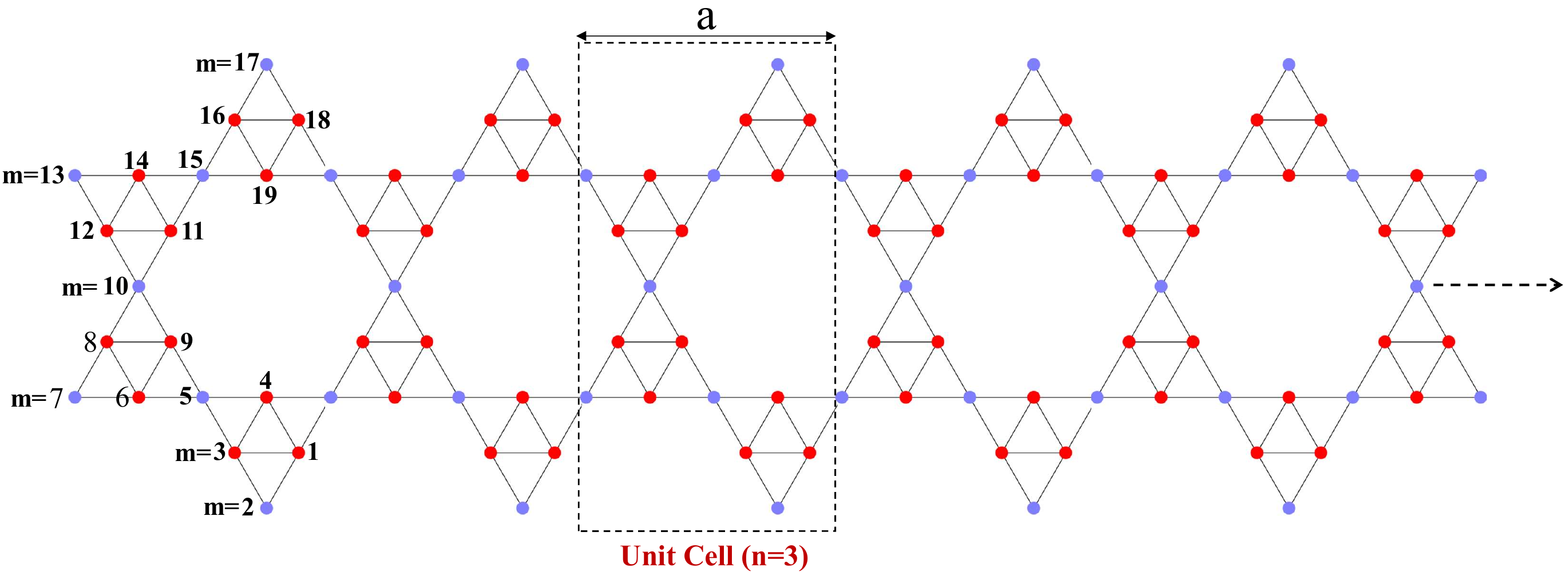}
\caption{(Color online)
Schematic illustration of the optimized geometry structure of the 1D boron kagome lattice used in our calculations. The dashed rectangle denotes the primitive translational unit cell for 1D-BTKL and $a$ is the length of the unit cell. Different colors of sublattice represent
distinct staggered potentials. The blue and red balls denote the B atoms sharing large and small triangles in the kagome lattice, respectively. For the sake of simplicity, for showing the various impurity positions, each boron atom in its unit cell is labeled with a number (m). The length of the lattice strip is determined by $L=Na$, with $N$,the number of unit cells.}
\label{schem1}
\end{figure*}

\begin{figure}
\includegraphics[width=7cm]{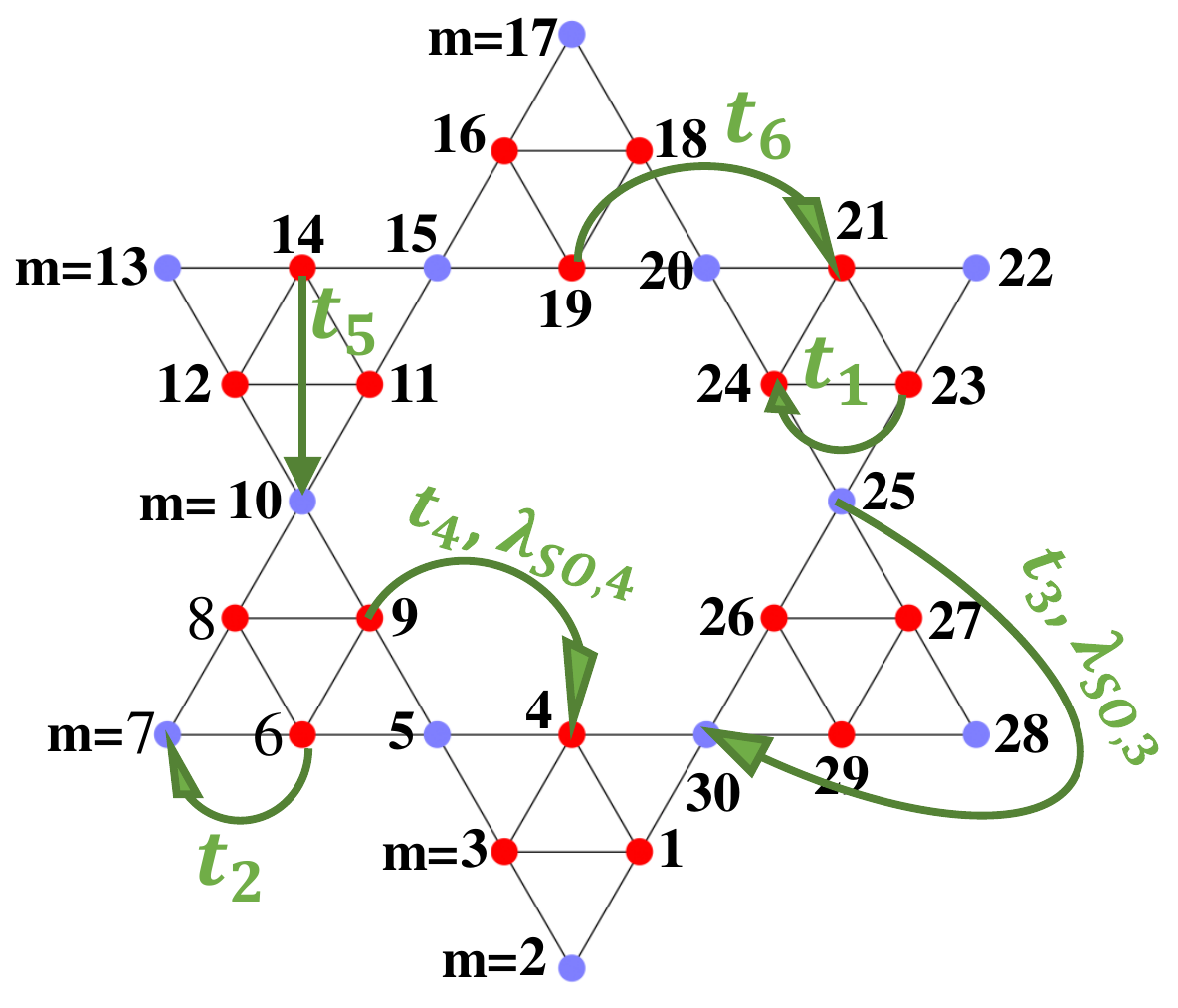}
\caption{(Color online)
Schematic view of the optimized geometry structure of the zero-dimensional (quantum dot) BTKL. The hopping parameters and spin-orbit couplings (SOC) used in the tight-binding Hamiltonian are shown by the green vectors. For the sake of simplicity, each boron atom is labeled with a number (m). }
\label{schem2}
\end{figure}

The most energetically stable form of monolayer BTKL~\cite{Yu.Zhao}, B$_{9}$-KL, is shown in Fig.\ref{schem1}. A quantum dot form of BTKL is also presented in Fig.\ref{schem2}. As can be seen, the BTKLs contain triangles in triangles, a space group of $P6mm$(No.17), belong to the general kagome system with the subnet 2, in which each triangle of the kagome arrangement contains a stack of four triangles~\cite{Ziff2009,Yu.Zhao}. The blue and red balls denote the B atoms sharing large and small triangles in the kagome lattice, respectively.
For simplicity, in the 1D-BTKL each atom is labeled with a set $(n, m)$, where $m$ represents the atom number in each unit cell and $n$ denotes the unit cell number to which the $n$th atom belongs. The number of unit cells ($N$) is used to indicate the length of 1D-BTKL.
The cell number to which this atom belongs.

A strong research trend of the new millennium, is the use of realistic and precise models such as highly accurate parameterized tight binding model to understand the share electronic and magnetic properties of materials.
From the analysis of symmetry and orbital characters of the wave functions in a BTKL, by using maximally localized Wannier functions (MLWFs), an effective tight-binding Hamiltonian has been derived as~\cite{Yu.Zhao}

\begin{equation}\label{eqn:TBhamiltonian}
\mathcal{H}_0=  \sum_{\langle iE,jE \rangle} t_{iE,jE  } c_{iE}^{\dagger} c_{jE}  + \sum_{\langle iE,jE \rangle} t_{ iE,jC  } c_{iE}^{\dagger} c_{jC} +H.C.
\end{equation}

where $\langle iE,jE \rangle$ denotes nearest-neighbor pairs between the edge B atoms, $\langle iE,jC \rangle$ stands for
nearest-neighbor pairs between the edge and corner B atoms, and $c_{i}^{\dagger}$ ($c_{j}$) represents the creation (annihilation) operator of electrons at site $i (j)$. By using maximally localized Wannier functions (MLWFs), the hopping parameters obtained as $t_{1}=-2.31$ eV, $t_{2}=-2.10$ eV, $t_{3}=-0.08$ eV, $t_{4}=0.29$ eV, $t_{5}=0.52$ eV, and $t_{6}=0.39$ eV (see Fig.\ref{schem2}).

In this effective Hamiltonian, the spin-orbit couplings (SOC) are also obtained with the parameters $\lambda_\mathrm{SO,3}$ = 0.004 eV and $\lambda_\mathrm{SO,4}$ = 0.015 eV. It is worth noting that the SOCs could be enhanced by various methods, such as hydrogenation\cite{Balakrishnan2013}, introduction of transition metal adatoms~\cite{Weeks2011}, and substrate proximity effects\cite{Calleja2015}, which have been used for graphene.

In the case of a BTKL in the presence of staggered sublattice potential, the spin-independent staggered potential term is given by%
\begin{equation}
\mathcal{H}_{St}=\sum_{i\in C}\Delta_{St}c_{i,C}^{\dagger}c_{i,C}-%
\sum_{i\in E}\Delta_{St}c_{i,E}^{\dagger}c_{i,E}\text{,}%
\end{equation}
where $\Delta_{St}$ is the magnitude of the spin-independent staggered potential. Then the complete Hamiltonian of a 1D-BTKL exposed to an external staggering potential is given by

\begin{equation}\label{eqn:TBhamiltonian2}
\mathcal{H}=\mathcal{H}_0+\mathcal{H}_{St}
\end{equation}

After the Fourier transformation (owing to the translational invariant along the ribbon direction, $x$), we obtain the Hamiltonian in the momentum space,

\begin{equation}\label{eq:H-k}
\mathcal{H}_k=\sum_{\mathbf{k}}\Psi_{\mathbf{k}}^{\dagger}H(\mathbf{k}%
)\Psi_{\mathbf{k}}\text{,}%
\end{equation}
where in the basis $\Psi_{\mathbf{k}}=(c_{B_1,\mathbf{k}},c_{B_2,\mathbf{k}},c_{B_3,\mathbf{k}},\dots ,c_{B_p,\mathbf{k}}$), with $p$, the total number of B atoms in the unit cell of the BTKL sample, the Hamiltonian matrix is given by

\begin{equation}\label{eq:H-k1}
H(\mathbf{k})=H_{00}+H_{01}e^{-i k_x
a}+H_{01}^\dagger e^{i k_x a}
\end{equation}
in which $a$ is the unit-cell width and $H_{00}$ and $H_{01}$ describe coupling within the principal unit cell (intra-unit cell) and between the adjacent principal unit cells (inter-unit cell), respectively based on the real space tight binding model given by Eq.\ref{eqn:TBhamiltonian2}.

Furthermore, to understand the effects of the impurity position on the magnetic ground state, we study the local density of state (LDOS) of both 0D- and 1D-BTKL. Here the LDOS for site $i-$th, at a given position ${\bf r}$ and energy $E$, is computed numerically from

\begin{equation}\label{eq:rau}
\rho({\bf r},E)=\sum_{n }{|\psi_{n}({\bf r})|^2 \delta(E-E_{n })}
\end{equation}

where $n$ is the band index.

\begin{figure}
\includegraphics[width=1\linewidth]{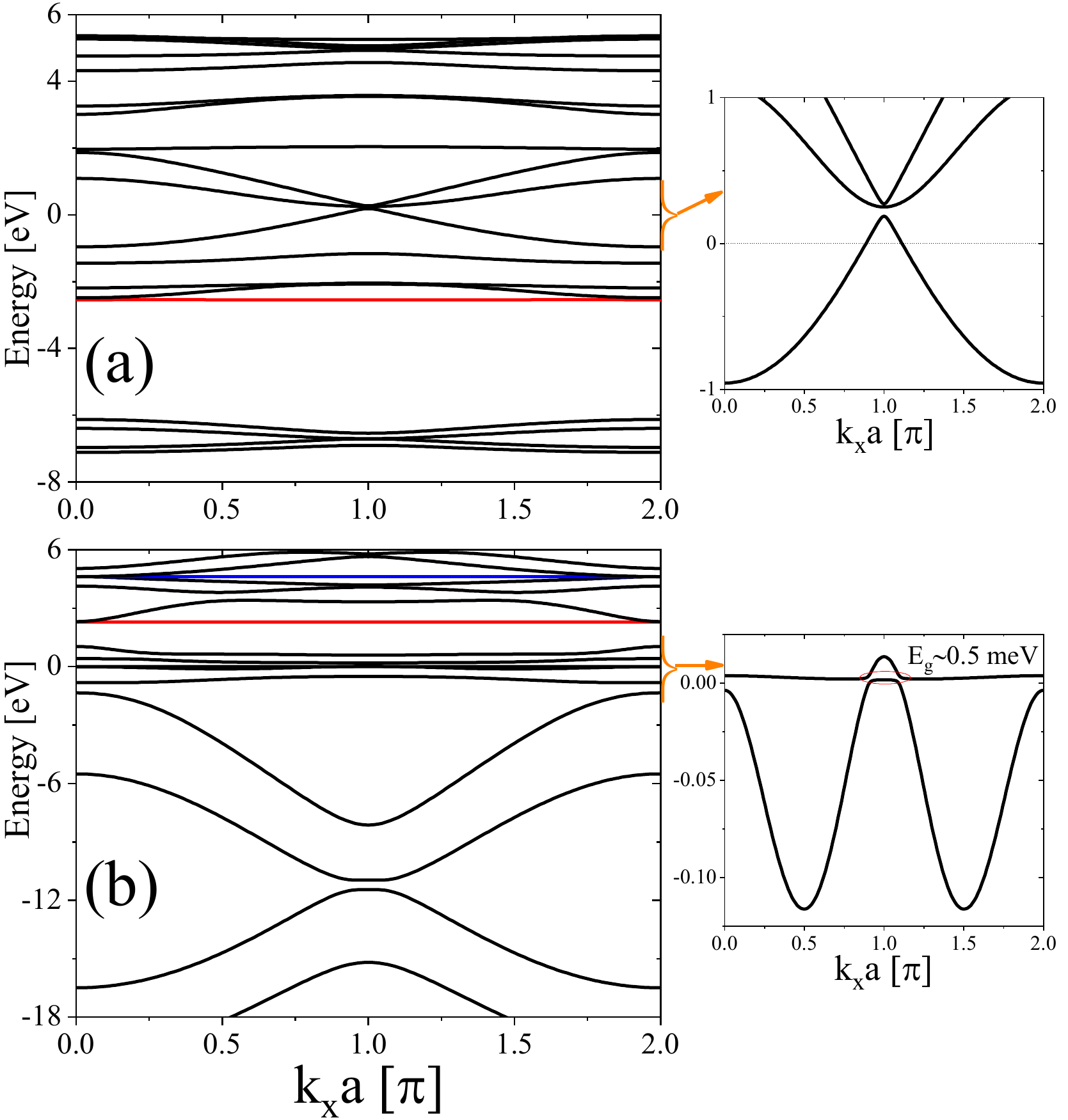}
\caption{(Color online)
The energy band structure of the 1D BTKLs with infinite lengths and $\lambda_\mathrm{SO,3}$ = 0.004 eV and $\lambda_\mathrm{SO,4}$ = 0.015 eV. In the bottom panels (b) we set all $t_i=t_1$. The flat bands are shown by red curves. The small right panels show zoom images of the areas as indicated by the yellow brackets on the large left panels.}
\label{DisK1}
\end{figure}

\begin{figure}
\includegraphics[width=1.02\linewidth]{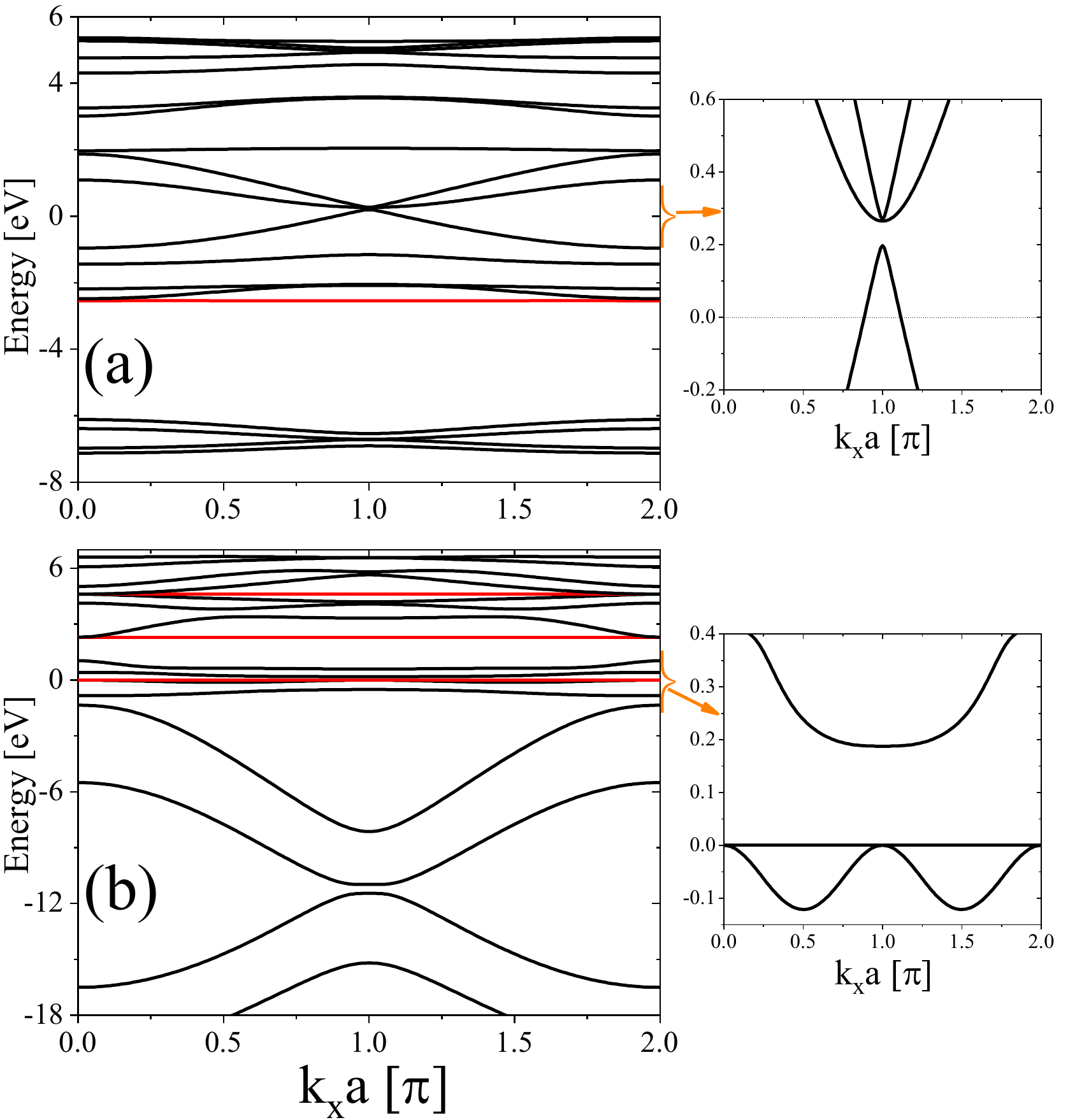}
\caption{(Color online)
The energy band structure of the 1D-BTKLs with infinite lengths and $\lambda_\mathrm{SO,3}$ = 0 and $\lambda_\mathrm{SO,4}$ = 0. In the bottom panels (b) we set all $t_i=t_1$. The complete and nearly flat bands are shown by red and blue curves, respectively. The small right panels show zoom images of the areas as indicated by the yellow brackets on the large left panels.}
\label{DisK2}
\end{figure}

The band structures of the 1D-BTKLs with infinite lengths at zero temperature, obtained by numerical diagonalization of the momentum space Hamiltonian Eq.\ref{eq:H-k1}, are plotted in Figs. \ref{DisK1} and \ref{DisK2}. As we said earlier, on the one hand, the topological features are affected by the SOCs and on the other hand the SOCs could be changed by various approaches, thus in order to clarify their effects on the band structure and exchange coupling we plot the band spectra for different hopping parameters $t_i$ and SOCs $\lambda_\mathrm{SO,3}$ and $\lambda_\mathrm{SO,4}$.

Figs. \ref{DisK1} and \ref{DisK2} show the band structures of 1D-BTKLs with $\lambda_\mathrm{SO,3}$ = 0.004, $\lambda_\mathrm{SO,4}$ = 0.015 eV and $\lambda_\mathrm{SO,3}=0$, $\lambda_\mathrm{SO,4}=0$, respectively. As an interesting consequence for the localized states, some flat bands with constant $E_n(k)$ and completely quenched kinetic energy, appear in the band spectra of the 1D-BTKLs. In both panels (b) we set all $t_i=t_1$ and in both panels (a) the hopping parameters $t_i$ are equal to the original values of the TB Hamiltonian, as shown in Fig.\ref{schem2}. The complete and nearly flat bands are shown by red and blue curves.
As we see from Figs. \ref{DisK1}(a) and \ref{DisK2}(a), as the zero Fermi energy crosses a band curve the 1D-BTKLs with the original hopping parameters of the Hamiltonian have a metallic behaviour. Furthermore, when we set all $t_i=t_1$, an energy gap around the zero Fermi energy appear for the case of $\lambda_\mathrm{SO,3}$ = 0.004,  $\lambda_\mathrm{SO,4}$ = 0.015 eV (see Fig. \ref{DisK1}(b)).

What the Fig.~\ref{DisK2} shows is the same as Fig.~\ref{DisK1} but for 1D-BTKLs with zero SOCs ($\lambda_\mathrm{SO,3}=\lambda_\mathrm{SO,4}=0$). Interestingly, three flat bands appear in the band structure of 1D-BTKLs by setting all $t_i=t_1$ (an isotropic TKL, panel (b)). One of these flat bands exactly lies at zero Fermi energy and two other ones are far from the zero energy point at energies about $E=2.43, 4.62$ eV. It is worth noting that the flat zero-energy band in momentum space touches a dispersive band at $k_xa=0,\pi, 2\pi$ (see Fig. \ref{DisK1}(b)). The another two flat bands touche the dispersive bands at $k_xa=0, 2\pi$.
This feature, already reported in several studies dedicated especially to classification of the flat bands in kagome-like lattices~\cite{BergmanPRB2008,MizoguchiPRB2019,RhimPRB2019}.

These flat bands can be generally categorized into two major classes, i.e., localized states and itinerant states with destructive interference.
It has long been known that, if the bandwidth is smaller than the band gap, then the high temperature fractional quantum Hall effect can appear for
partial filling of the flat band~\cite{Tang2011,Sarma2011,Mudry2011}. Thus, because of the vanishing bandwidth of these flat bands one can realize the high temperature fractional quantum Hall state in the 1D-BTKLs.

For a kagome lattice system with an isotropic nearest neighbor hopping integral~\cite{Sarma2007}, there exist one flat and two dispersive bands. The latter two have a similar form to those in graphene, and touch at two inequivalent Dirac points~\cite{YuS-L}.
It is shown that depending on the presence or absence of the discontinuity of the corresponding eigenfunctions, a band touching can be identified to be singular or nonsingular~\cite{RhimPRB2019}.
From these figures we find that the flat band energies can be tuned by changing the hopping parameters and SOCs. Moreover, due to the electronic band gap the 1D-BTKLs may be the next candidate for application in optoelectronic devices.

\subsection{Numerical results for the RKKY interaction}\label{sec:Numer-R}

In this section, we present in the following our main numerical results for the comprehensive evaluation of the RKKY exchange (Eq.\ref{chiE}) in the zero and one-dimensional boron triangular kagome lattices.

\begin{centering}
\begin{figure}
\includegraphics[width=1.0\linewidth]{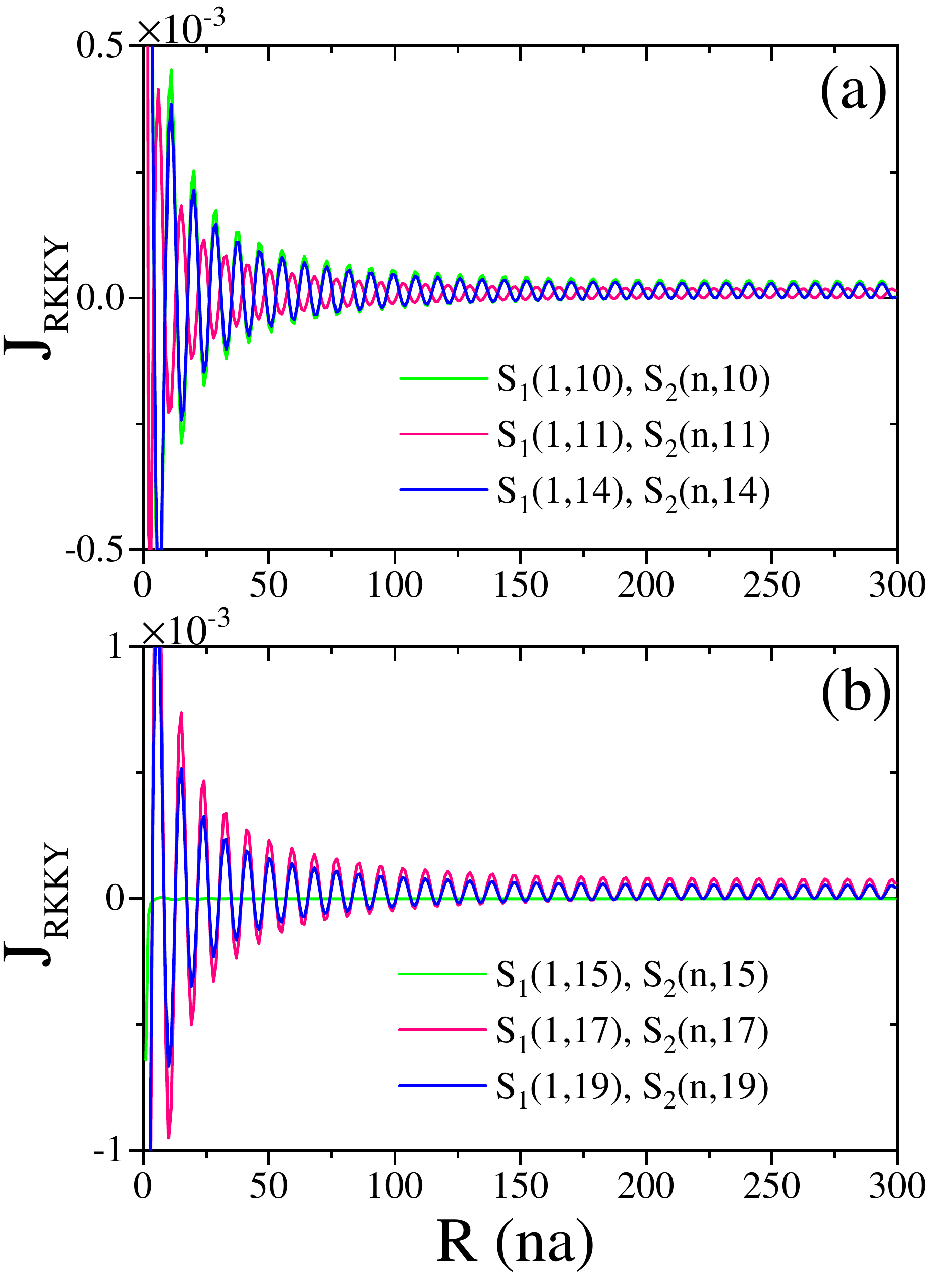}
\caption{(Color online) The range function of RKKY interaction as a function of the distance between localized spins (in units of the unit cell length), for the intrinsic case ($E_F=0$). The first impurity is located at the first unit cell at different lattice sites with coordinate $(1,m)$, with $m=10,11,14$ in panel(a) and $m=15,17,19$ in panel(b), as shown in Fig.\ref{schem1}, and the second one moves at lattice sites $(n,m)$.}
\label{chi_R0}
\end{figure}
\end{centering}

Figure \ref{chi_R0} shows the effective exchange interaction as a function of the distance between localized spins (in units of the unit cell length), for the intrinsic case ($E_F=0$) for different spatial configurations of the magnetic impurities. The first impurity is located at the first unit cell (an edge unit cell) at lattice sites with coordinate $(1,m)$ with $m=10,11,14$ in panel (a) and $m=15,17,19$ in panel (b), as shown in Fig.\ref{schem1}, and the second one is moved at lattice sites $(n,m)$.
It is clear that the RKKY interaction changes sign as a function of distance.
For the case when both spins are at the lattice sites with  $(n,15)$ (the panel (b)-green curve) the result is quite different, because in this situation the RKKY interaction is at least three order of magnitude smaller than the other configurations. As indicated, in this case the RKKY coupling shows an oscillatory behavior in R and decays fast with a short-ranged behavior.

\begin{centering}
\begin{figure}
\includegraphics[width=1.0\linewidth]{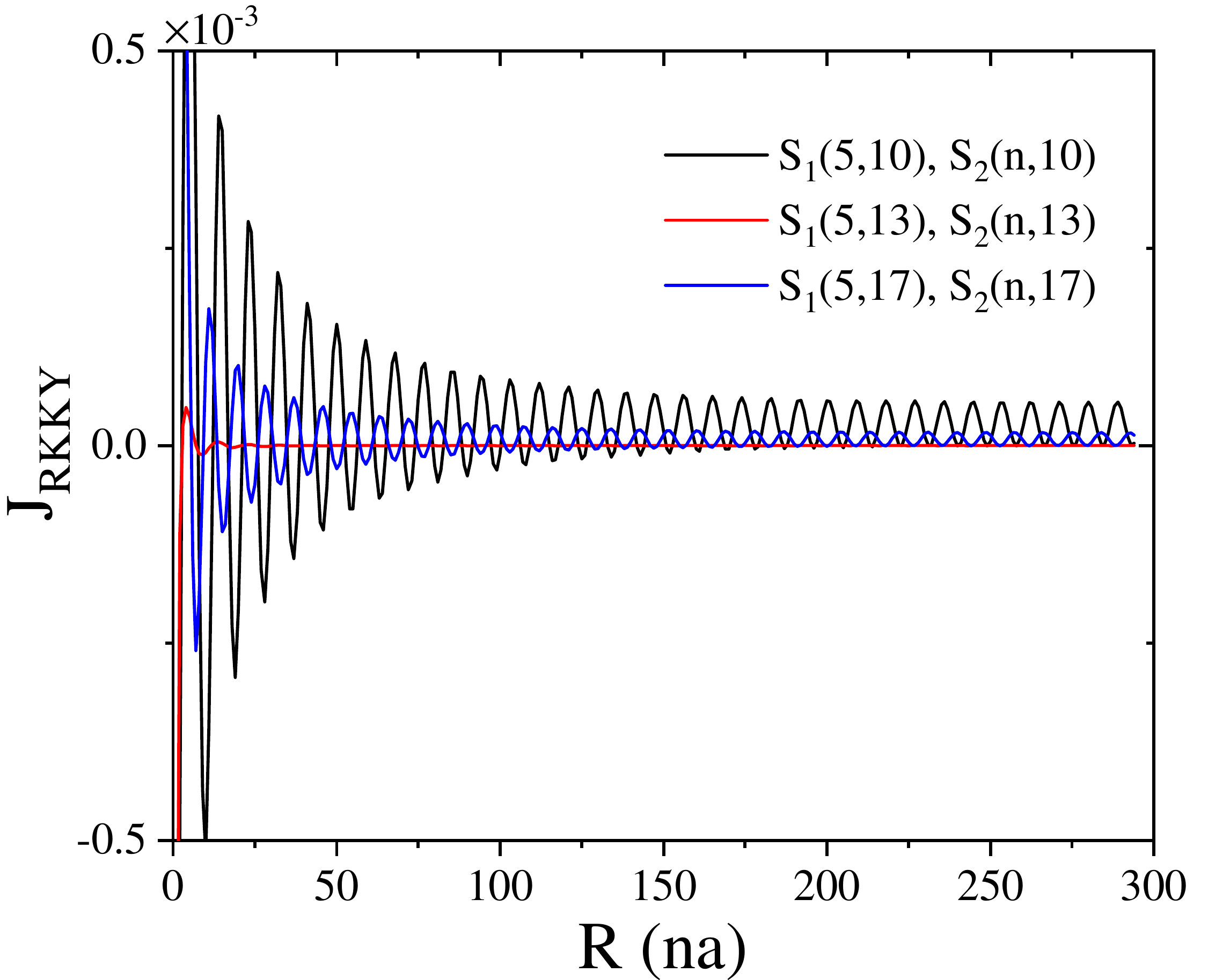}
\caption{(Color online) The same as Fig.\ref{chi_R0}, but here, the first impurity is located at the fifth unit cell at lattice sites with coordinate $(5,m)$ with $m=10,13,17$ , as shown in Fig.\ref{schem1}, and the second one moves at lattice sites $(n,m)$ with $n=6,7,8,9,....$.}
\label{chi_R1}
\end{figure}
\end{centering}

Figure \ref{chi_R1} is the same as Fig. \ref{chi_R0}, but here, the first impurity is located at the fifth unit cell (an bulk unit cell) at different lattice sites with coordinate $(5,m)$ with $m=10,13,17$ , as shown in Fig.\ref{schem1}, and the second one moves at lattice sites $(n,m)$ with $n=6,7,8,9,....$. In analogy to the case of both spins are at the lattice sites with $(n,15)$, the RKKY coupling shows an oscillatory behavior in R and decays fast with a short-ranged behavior, when both spins are inside the 1D-TKL at the lattice sites with $(n,13)$. This is reasonable because both sites $m=13$ and $m=15$.

\begin{centering}
\begin{figure}
\includegraphics[width=1.0\linewidth]{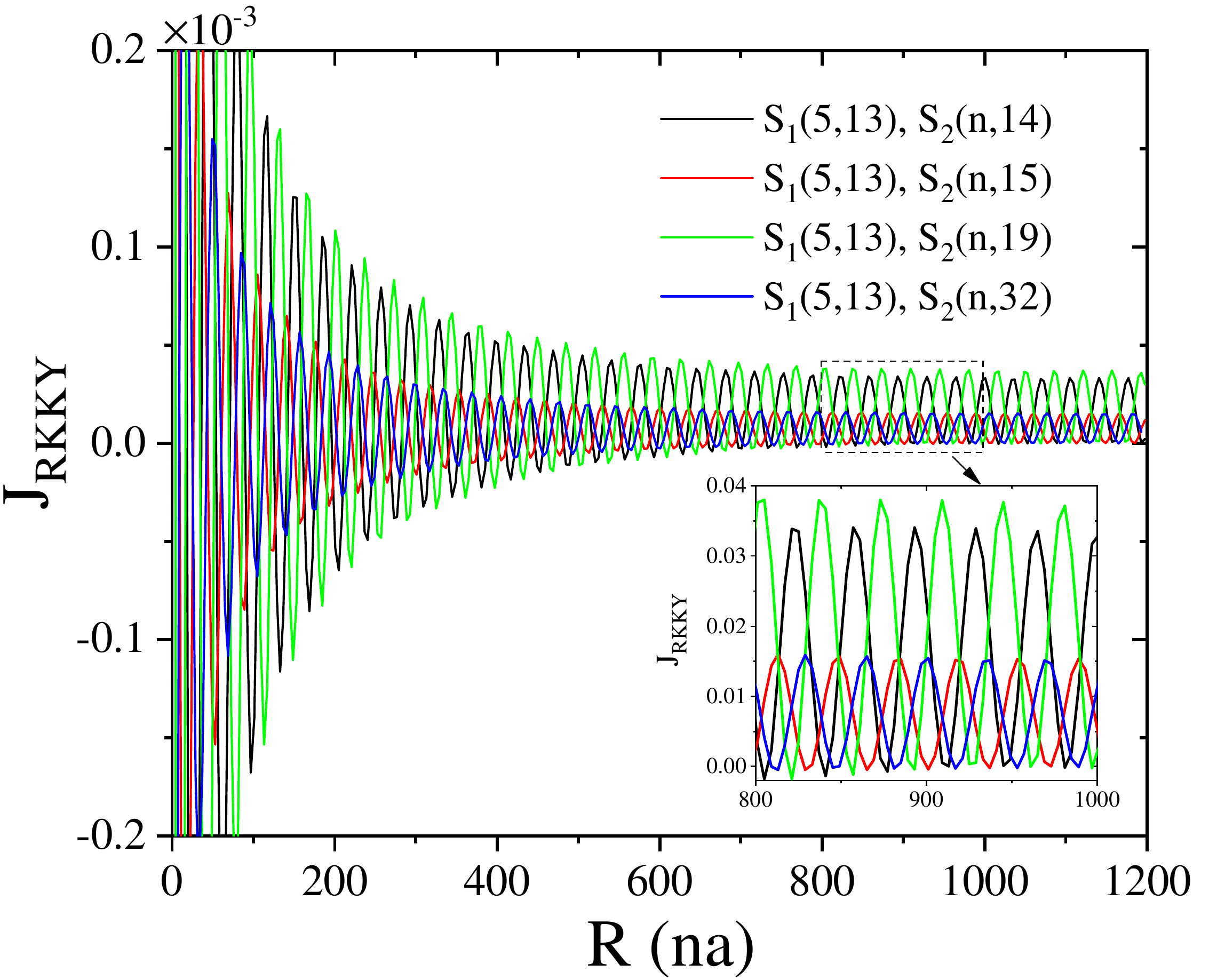}
\caption{(Color online) The same as Fig.\ref{chi_R1}, but here, the first impurity is located at the fifth unit cell at different lattice sites with coordinate $(5,13)$, and the second one moves at lattice sites $(n,m)$ with $m=14,15,19,32$ and $n=6,7,8,9,....$.}
\label{chi_R2}
\end{figure}
\end{centering}

What the Fig. \ref{chi_R2} shows is the same as Fig. \ref{chi_R1} but here, the first impurity is located at lattice site with coordinate $(5,13)$ (in the fifth unit cell, a bulk unit cell ) with $m=10,13,17$ and the second one moves at lattice sites $(n,m)$ with $n=6,7,8,9,....$ and $m=14,15,19,32$.
It is shown that the RKKY interaction has an oscillating behaviour in terms of the distance between localized spins, and more importantly, both sign and strength of the interactions can be tuned by the impurity positions.

We will now focus on the qualitative behavior of the RKKY interaction as function of the Fermi energy.

\begin{centering}
\begin{figure*}
\includegraphics[width=1.0\linewidth]{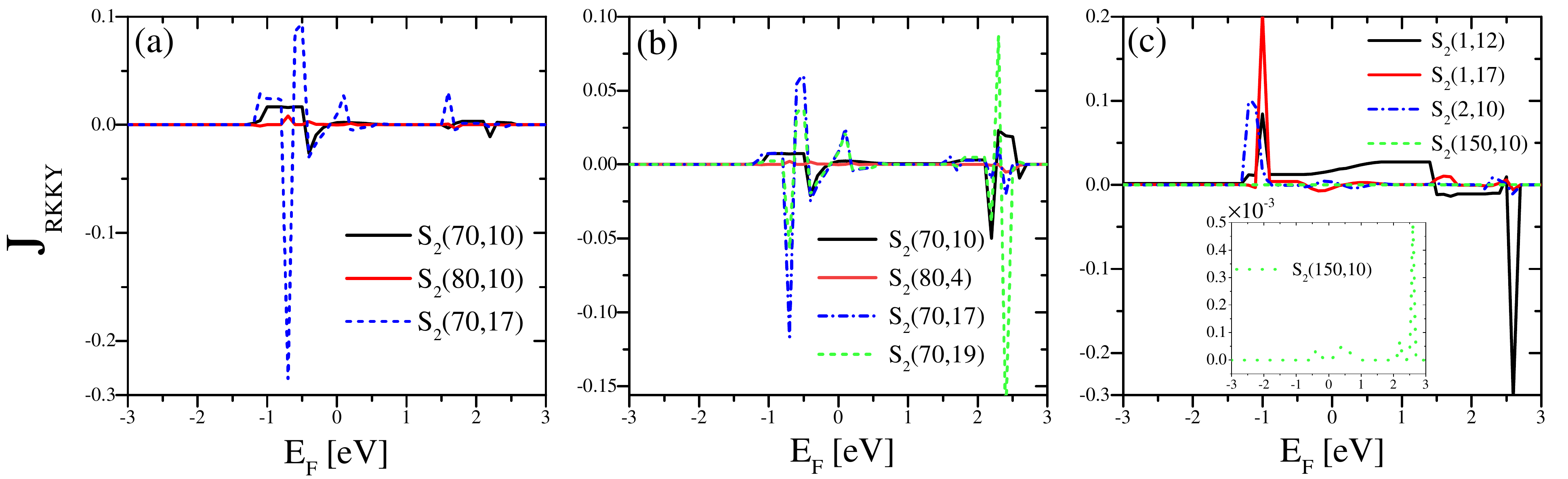}
\caption{(Color online) The variation of the RKKY coupling versus the Fermi energy for 1D-TKLs with $N=150$, for different spatial configurations of the magnetic impurities. The first impurity is located at a lattice site with coordinate $(70,2)$ in panel (a), $(70,4)$ in panel (b) and $(1,10)$ in panel (c).}
\label{chi1D_EF1}
\end{figure*}
\end{centering}

Figure \ref{chi1D_EF1} displays the variation of the RKKY coupling versus the Fermi energy for 1D-TKLs with $N=150$, for different spatial configurations of the magnetic impurities. The first impurity is located at a lattice site with coordinate $(70,2)$ in panel (a), $(70,4)$ in panel (b) and $(1,10)$ in panel (c).

\begin{centering}
\begin{figure*}
\includegraphics[width=1.0\linewidth]{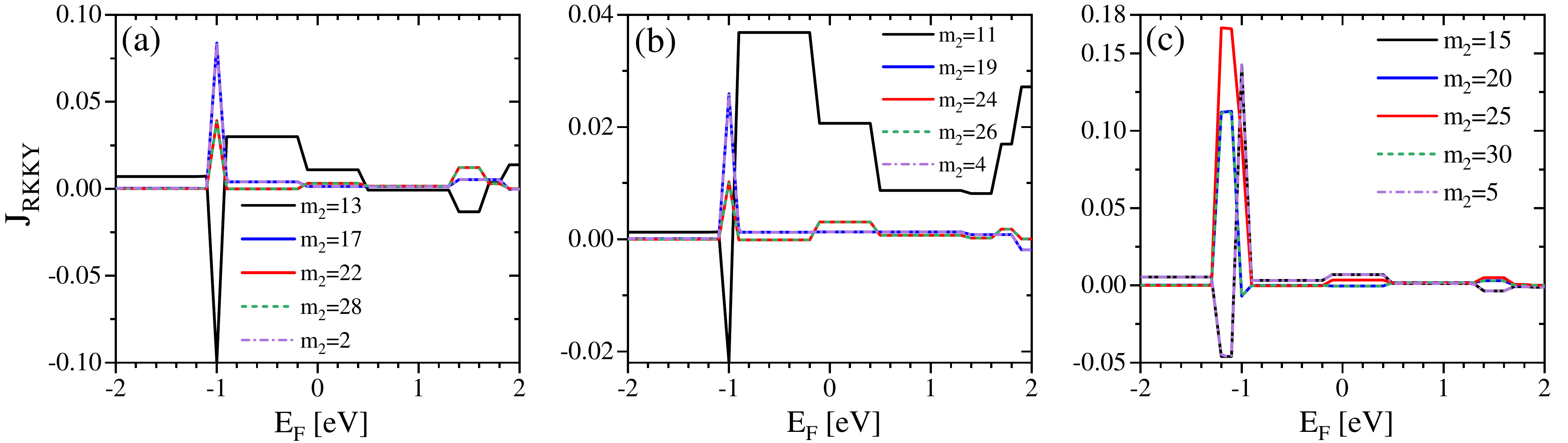}
\caption{(Color online) The variation of the RKKY coupling versus the Fermi energy for 0D-TKL (TKL quantum dot), for different spatial configurations of the magnetic impurities. The first impurity is located at the lattice site with $m=1$ in all panels and the second one is located at the lattice sites with $m=2,13,17,22,28$ in panel (a) and $m=4,11,19,24,26$ in panel (b) and $m=5,15,20,25,30$ in panel (c).}
\label{chi1K_EF1}
\end{figure*}
\end{centering}

Figure \ref{chi1K_EF1} shows the variation of the RKKY exchange parameter $J_{RKKY}$, with respect to the Fermi energy for the 0D-TKL (TKL quantum dot).
The first impurity is located at the lattice site with $m=1$ in all panels and the second one is located at the lattice sites with $m=2,13,17,22,28$ in panel (a) and $m=4,11,19,24,26$ in panel (b) and $m=5,15,20,25,30$ in panel (c).

\begin{centering}
\begin{figure}
\includegraphics[width=1.0\linewidth]{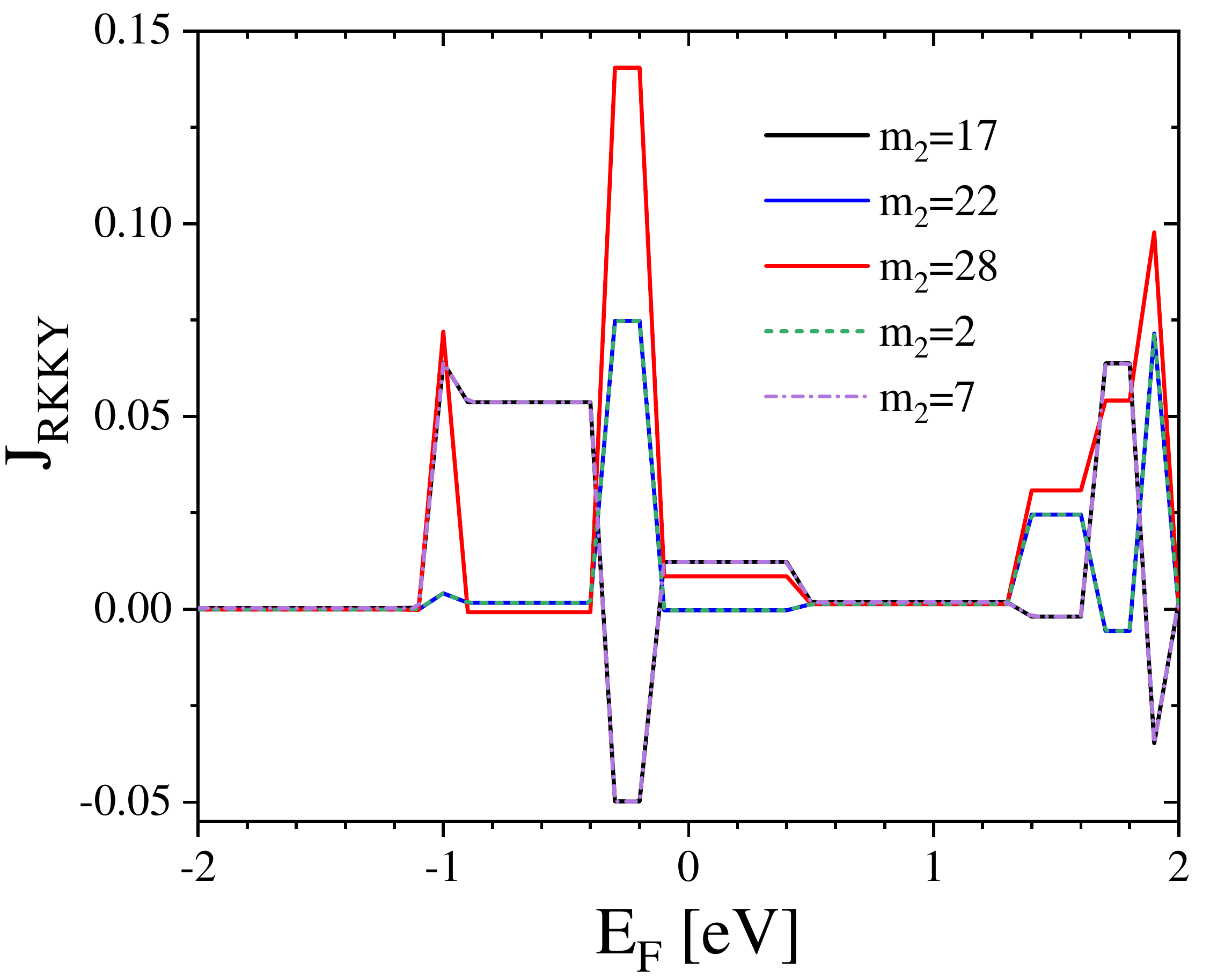}
\caption{(Color online) The same as Fig.~\ref{chi1K_EF1}, but here, the first impurity is located at the lattice site with $m=3$ and the second one at sites $m=2,7,17,22,28$.}
\label{chi1K_EF2}
\end{figure}
\end{centering}

What the Fig. \ref{chi1K_EF2} displays is the same as Fig. \ref{chi1K_EF1}, for a 0D-TKL but here the first impurity is located at the lattice site with $m=3$ and the second one at sites $m=2,7,17,22,28$.

The importance of the 2D Heisenberg-like of the RKKY interaction in the TKLs, as shown in Figs. \ref{chi1D_EF1}, \ref{chi1K_EF1} and \ref{chi1K_EF2}, is recognized as the appearance of the magnetization plateau in the exchange profile of the TKLs versus the Fermi energy.
Such a long-range interaction is intuitively expected to arise from the RKKY interactions in the rare-earth tetraborides family such as $RB_4$ ~\cite{Siemensmeyer08} and $TbB_4$~\cite{YoshiiPRL2008}.
However, the RKKY plateau has not been reported before, to the best of our knowledge.
One of the pioneering theoretical works where the appearance of the magnetization plateau was predicted is the paper of Hida~\cite{Hida}.

As seen in these figures, the value of the magnetization plateaus varies between different phases, namely ferromagnetic and antiferromagnetic orders.
Similarly, looking at the finite-size plateaux (see Figs. \ref{chi1K_EF1} and \ref{chi1K_EF2}), it seems that finite-size effects are strong both
for the width and the location of these plateaux.

Importantly, the width and location of the plateaus show tunability in magnetic RKKY coupling on the Fermi energy and spatial configurations of the impurities in both 0D and 1D-TKL. This proves to be an alternative approach to tuning the magnetic ground state in TKLs.

\begin{centering}
\begin{figure*}
\includegraphics[width=1.0\linewidth]{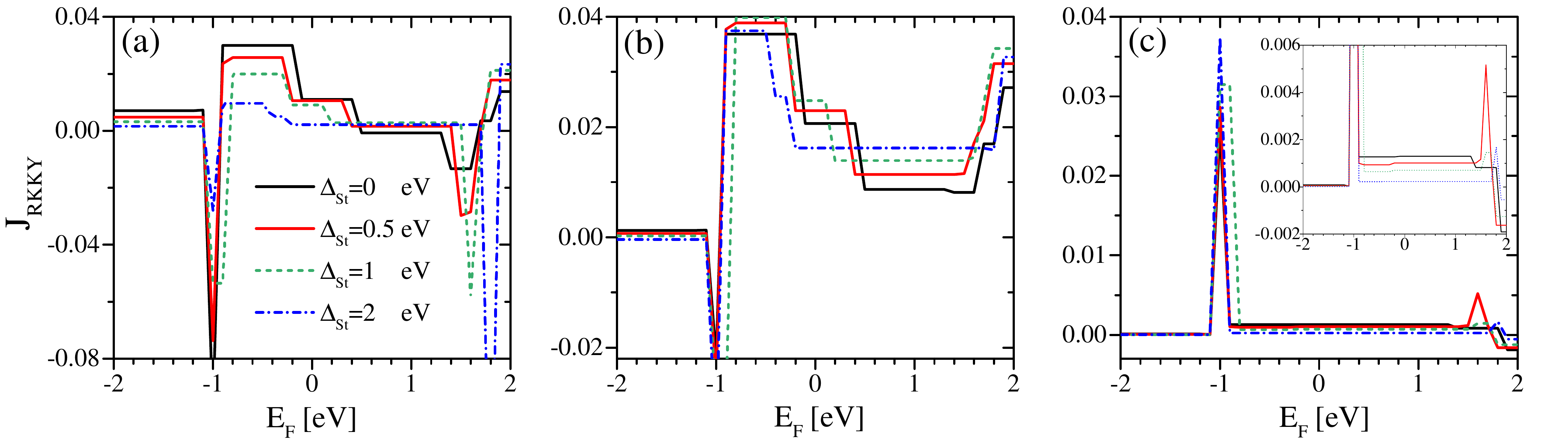}
\caption{(Color online) The variation of $\chi$ versus the Fermi energy for 0D-TKL (TKL quantum dot), for several fixed values of the sublattice staggered potential $\Delta_{\text St}$. In all panels the first impurity is located at the lattice site with $m=1$ and the second one is located at the lattice site $m=3$ in panel (a) and $m=5$ in panel (b) and $m=25$ in panel (c). The inset in panel (c) shows a zoom view around the zero RKKY. }
\label{chi1K_EF3}
\end{figure*}
\end{centering}

Focusing on the expected changes of the plateaus from the sublattice staggered potential, we turn now to investigate the tuning of the width of the RKKY plateaus using an external staggered potential in 0D-TKL. We show the width of the RKKY plateaus for different value of the sublattice staggered potential $\Delta_{\text St}$ in Fig.~\ref{chi1K_EF3} as a function of the Fermi energy. In all cases, the first impurity is fixed at a given initial position with $m_1=1$ and the second one is fixed at high symmetry coordinates, i.e., $m_2=3$ in panel (a) and $m_2=5$ in panel (b) and $m_2=25$ in panel (c). As shown in Fig.~\ref{chi1K_EF3}, it is striking that the width of the RKKY plateaus are highly tunable with respect to the external staggered potential and the Fermi energy as well as the spatial configurations of the magnetic impurities.
Indeed, the most remarkable observation in this paper is the potential and Fermi energy variation of both width and location of the RKKY plateaus for both 0D- and 1D-TKLs, as shown in Figs. ~\ref{chi1K_EF1}, ~\ref{chi1K_EF2} and ~\ref{chi1K_EF3}.

In addition, understanding the sublattice-dependent of local density of states (LDOS) is essential to assess the configuration-dependent exchange interaction. To do so, it is necessary to obtain the diagonal components of the unperturbed Green’s function matrix $ G({\bf r,r},E)$, for a lattice site at position $ {\bf r}$ and energy $E$.

\begin{centering}
\begin{figure}
\includegraphics[width=1.0\linewidth]{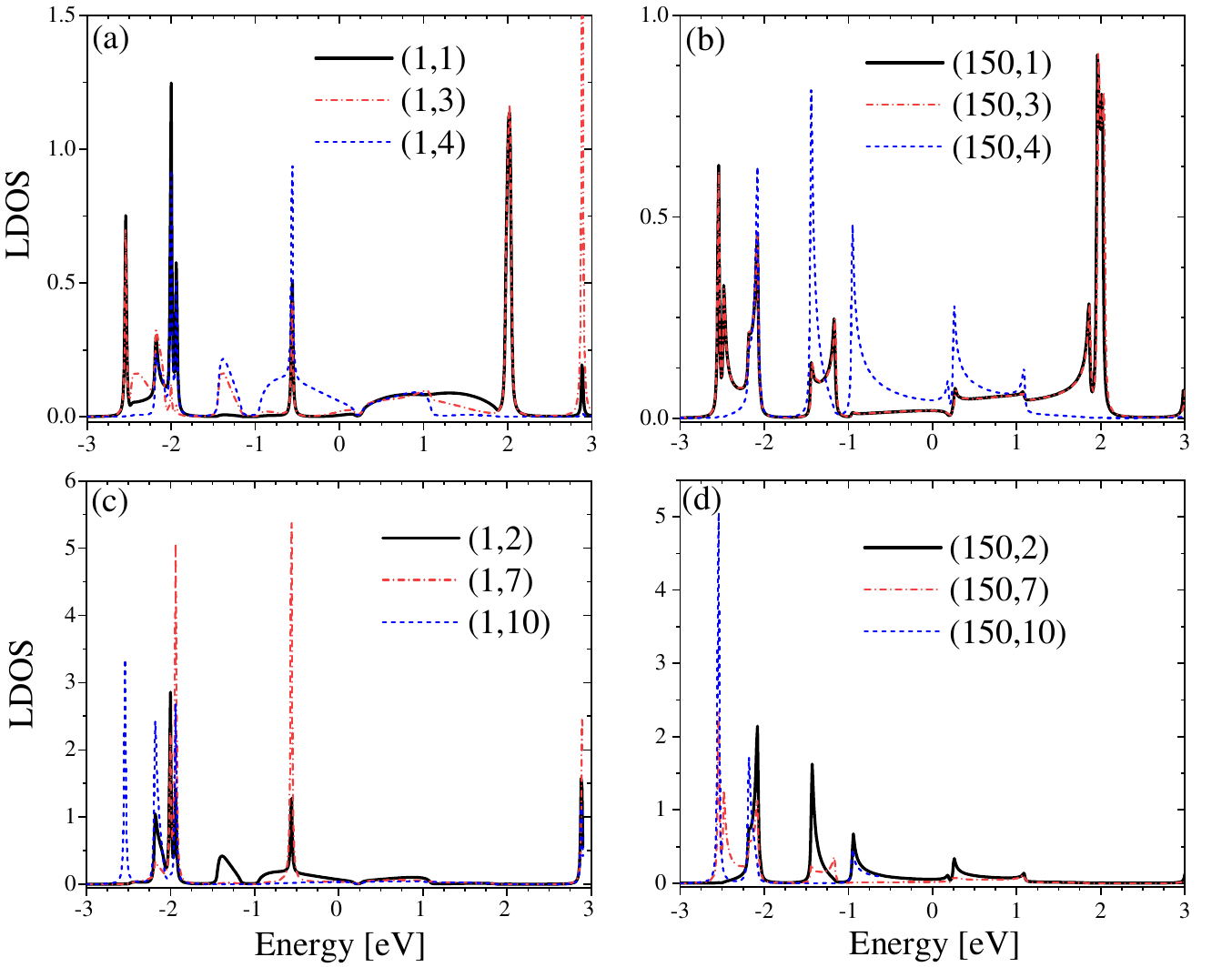}
\caption{(Color online) Calculated local density of states for a 1D-TKL with $N = 300$, for different lattice sites (for both edge and bulk sites).}
\label{LDOS1DKL}
\end{figure}
\end{centering}

\begin{centering}
\begin{figure}
\includegraphics[width=1.0\linewidth]{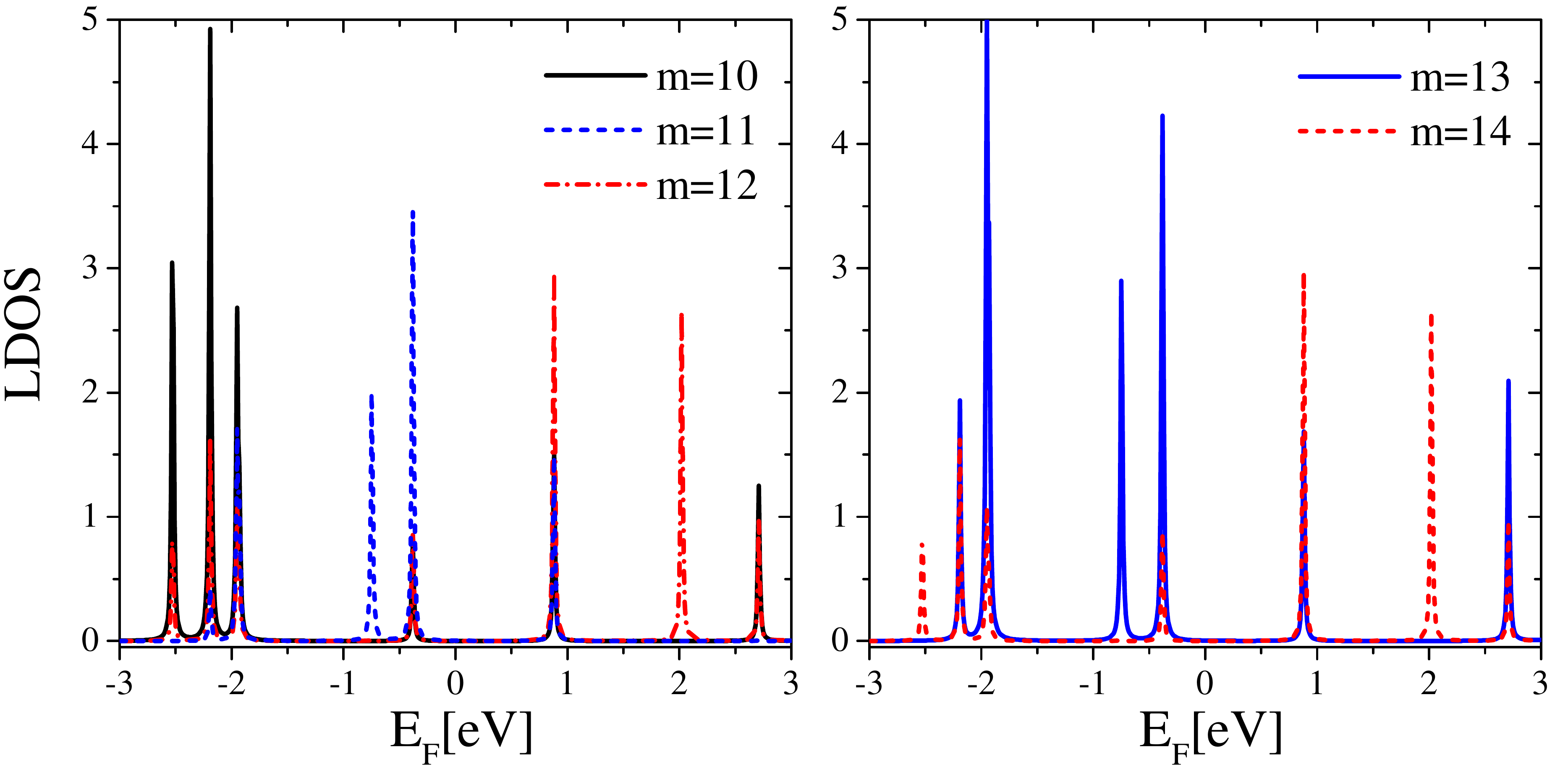}
\caption{(Color online) Calculated local density of states for a 0D-TKL, for different lattice sites.  }
\label{LDOS1K}
\end{figure}
\end{centering}

Fig.~\ref{LDOS1DKL} illustrates the LDOS for a 1D-TKL with $N=300$, for different lattice sites with coordinate $m$ (for both edge and bulk sites) and the Fig.~\ref{LDOS1K} shows the LDOS for a 0D-TKL, for different lattice sites $m$.
Specifically, for the 0D-TKLs the LDOS peaks become much sharper than that of the 1D model.
Looking at the LDOS corresponding to the various lattice sites, it is clear that the higher values of LDOS appear on atoms at the corners of the large triangles in the kagome lattice (the blue atoms in Fig.~\ref{schem2}).

\section{summary}\label{sec:summary}
In summary, we have implemented a tight-binding method to study the electronic properties and magnetic interactions in both zero- and one-dimensional triangular Kagome lattice models.
We firstly study the electronic properties of both 0D and 1D-TKL in the presence of staggered sublattice potential, then, by analyzing the  Ruderman-Kittel-Kasuya-Yoshida (RKKY) interaction in these lattice models, the magnetic ground states of both 0D- and 1D-TKL in the presence of two magnetic impurities are investigated, by using the real-space Green’s function approach.

It is found that the 1D channels of TKL show different electronic and magnetic behaviors due to different values of the hopping integrals and spin-orbit couplings. For an isotropic 1D-BTKL with zero SOCs, three flat bands appear in the band structure: one of these flat bands exactly lies at zero Fermi energy and two other ones are far from the zero energy point. It is worth noting that the flat zero-energy band in momentum space touches a dispersive band at $k_xa=0,\pi, 2\pi$. The another two flat bands touche the dispersive bands at $k_xa=0, 2\pi$.

Another important salient feature of the TKLs is the emergence of the RKKY plateaus versus the Fermi energy.
These RKKY plateaus have not been reported before, to the best of our knowledge.
The RKKY plateaus have been examined in detail with respect to the external staggered potential and the Fermi energy as well as the spatial configurations of the magnetic impurities.
The most remarkable observation is the potential and Fermi energy variation of the width and location of the RKKY plateaus in both 0D and 1D-TKLs.
The spatial configurations of the magnetic impurities can also dramatically change the quality and quantity of the RKKY plateaus.
Our results provide crucial guidance in designing further experiments to search for the realizing of the flat bands and magnetization plateau phases in spintronics and pseudospin electronics devices based on TKLs.

\end{document}